\documentstyle[11pt]{article}
\begin{document}
\title{\Large\bf  Regarding the axial-vector mesons}

\author{\small De-Min Li $^{1,2}$\footnote{E-mail: lidm@zzu.edu.cn},~~Bing Ma $^{1}$,~~Hong Yu $^{2}$\\
\small  $^1$ Department of Physics, Zhengzhou University,
Zhengzhou, Henan 450052, P. R. China\footnote{Mailing address}\\
\small  $^2$ Institute of High Energy Physics, Chinese Academy of
Sciences, Beijing 100039, P. R. China\\}
\date{\today}
\maketitle
\vspace{0.5cm}

\begin{abstract}

The implications of the $f_1(1285)-f_1(1420)$ mixing for the $K_1(^3P_1)-K_1(^1P_1)$ mixing angle
is investigated. Based on the $f_1(1285)-f_1(1420)$ mixing angle $\sim 50^\circ$ suggested from the
analysis for a substantial body of data concerning the $f_1(1420)$ and $f_1(1285)$, the masses of
the $K_1(^3P_1)$ and $K_1(^1P_1)$ are  determined to be $\sim 1307.35\pm 0.63$ MeV and $1370.03\pm
9.69$ MeV, respectively, which therefore suggests that the $K_1(^3P_1)-K_1(^1P_1)$ mixing angle is
about $\pm (59.55\pm 2.81)^\circ$.  Also, it is found that the mass of the $h^\prime_1(^1P_1)$
(mostly of $s\bar{s}$) state is about $1495.18\pm 8.82$ MeV. Comparison of the predicted results
and the available experimental information of the $h_1(1380)$ shows that without further
confirmation on the $h_1(1380)$, the assignment of the $h_1(1380)$ as the $s\bar{s}$ member of the
$^1P_1$ meson nonet may be premature.
\end{abstract}

\vspace{0.5cm}

{\bf Key words:} axial-vector mesons; hadron mass

{\bf PACS numbers:}14.40.Ev; 12.40.Yx

\newpage

\baselineskip 24pt

\section{Introduction}
\indent \vspace*{-1cm}

 The strange axial vector mesons provide interesting possibilities to study the QCD in the nonperturbative regime by
 the mixing of the $^3P_1$ and $^1P_1$ states. In the exact SU(3) limit, the $K_1(^3P_1)$ and
 $K_1(^1P_1)$ do not mix, just as the $a_1$ and $b_1$ mesons do not mix.
 For the strange quark mass
 greater than the up and down quark masses so that SU(3) is broken, also, the $K_1(^3P_1)$ and
 $K_1(^1P_1)$ do not possess definite C-parity, therefore these states can in principle mix
 to give the physical $K_1(1270)$ and $K_1(1400)$.

 In the literature,
 the mixing angle of the $K_1(^3P_1)$ and
 $K_1(^1P_1)$, $\theta_K$ has been estimated by some different approaches, however, there is not yet a
 consensus on the value of $\theta_K$.   As the optimum fit to the data as of
 1977, Carnegie et al. finds $\theta_K=(41\pm 4)^\circ$\cite{carnegie}. Within the heavy quark effective theory
Isgur and Wise predict two possible mixing angles, $\theta_K\sim 35.3^\circ$ and $\theta_K\sim
-54.7^\circ$\cite{isgur}. Based on the analysis of $\tau\rightarrow \nu K_1(1270))$ and
$\tau\rightarrow \nu K_1(1400))$, Rosner suggests $\theta_K\sim 62^\circ$\cite{rosner},  Asner et
al. gives $\theta_K= (69\pm 16\pm 19)^\circ$ or $(49\pm 16\pm 19)^\circ$\cite{asner}, and Cheng
obtains $\theta_K= \pm 37^\circ$ or $\pm 58^\circ$\cite{cheng}. From the experimental information
on masses and the partial rates of $K_1(1270)$ and $K_1(1400)$, Suzuki finds two possible solutions
with a two-fold ambiguity, $\theta_K\sim 33^\circ$ or $57^\circ$\cite{suzuki}. A constraint
$35^\circ\leq \theta_K \leq 55^\circ$ is predicted by Burakovsky et al. in a nonrelativistic
constituent quark model\cite{burakovksy}, and within the same model, the values of $\theta_K\simeq
(31\pm 4)^\circ$ and $\theta_K\simeq (37.3\pm 3.2)^\circ$ are suggested by Chliapnikov\cite{nrqm}
and Burakovsky\cite{prd57}, respectively. The calculations for the strong decays of $K_1(1270)$ and
$K_1(1400)$ in the $^3P_0$ decay model suggests $\theta_K\sim 45^\circ$\cite{blundell,barnes}. The
mixing angles $\theta_K\sim 34^\circ$\cite{isgod}, $\theta_K\sim 5^\circ$\cite{godkok} are also
presented within a relativized quark model. More recently, Vijande et al. suggests $\theta_K\sim
55.7^\circ$ based on the calculations in a constituent quark model\cite{vijande}.

It is widely believed that the $f_1(1285)$ and $f_1(1420)$ are the isoscalar states of the $^3P_1$
meson nonet\cite{pdg}. The analysis of the Gell-Mann-Okubo mass formula, SU(3) coupling formula,
radiative decay of the $f_1(1285)$, $\gamma\gamma^\ast$ decays of the $f_1(1285)$ and $f_1(1420)$,
and the radiative $J/\psi$ decays performed by Close and Kirk\cite{zphysc76} indicates that these
various data are independently consistent with the $f_1(1285)-f_1(1420)$ mixing angle $\alpha\sim
50^\circ$ (in the singlet-octet basis). This value of $\alpha\sim 50^\circ$ is also supported by
the calculations performed by \cite{vijande,cpl17,jpa35,epjc37}.

We shall show below that the mass of the $K_1(^3P_1)$ can be related to the mass matrix describing
the mixing of the $f_1(1285)$ and $f_1(1420)$, and the $f_1(1285)-f_1(1420)$ mixing angle can give
a constraint on the mixing of $K_1(^3P_1)-K_1(^1P_1)$. The main purpose of the present work is to
discuss the implications of the $f_1(1285)-f_1(1420)$ mixing for the $K_1(^3P_1)-K_1(^1P_1)$ mixing
angle.

\section{The mixing angle of the $K_1(^3P_1)$ and
 $K_1(^1P_1)$}
\indent\vspace*{-1cm}

In the $N=(u\bar{u}+d\bar{d})/\sqrt{2}$, $S=s\bar{s}$ basis, the mass-squared matrix describing the
mixing of the $f_1(1420)$ and $f_1(1285)$ can be written as\cite{jpg27}
\begin{equation}
M^2=\left(\begin{array}{cc}
M^2_{a_1(^3P_1)}+2\beta&\sqrt{2}\beta X\\
\sqrt{2}\beta X&2M^2_{K_1(^3P_1)}-M^2_{a_1(^3P_1)}+\beta X^2
\end{array}\right),
\label{mix}
\end{equation}
where $M_{a_1(^3P_1)}$ and $M_{K_1(^3P_1)}$ are the masses of the states $a_1(^3P_1)$ and
$K_1(^3P_1)$, respectively;
 $\beta$ denotes the total
annihilation strength of the $q\bar{q}$ pair for the light flavors $u$ and $d$; $X$ describes the
SU(3)-breaking ratio of the nonstrange and strange quark propagators via the constituent quark mass
ratio $m_u/m_s$. The masses of the two physical isoscalar states $f_1(1420)$ and $f_1(1285)$, $M_1$
and $M_2$, can be related to the matrix $M^2$ by the unitary matrix $U$
\begin{eqnarray}
M^2=U^\dagger\left(\begin{array}{cc}
M^2_1&0\\
0&M^2_2
\end{array}\right)U,
\label{dig}
\end{eqnarray}
and the physical states $f_1(1420)$ and $f_1(1285)$ can be expressed as
\begin{eqnarray}
\left(\begin{array}{l}
f_1(1420)\\
f_1(1285)
\end{array}\right)=
U\left(\begin{array}{l}
N\\
S
\end{array}\right).
\label{NS}
\end{eqnarray}

Also, in the basis $\mbox{\bf 8}=(u\bar{u}+d\bar{d}-2 s\bar{s})/\sqrt{6}$, $\mbox{\bf
1}=(u\bar{u}+d\bar{d}+s\bar{s})/\sqrt{3}$, the mixing of the $f_1(1420)$ and $f_1(1285)$ can
expressed by
\begin{eqnarray}
\left(\begin{array}{l}
f_1(1420)\\
f_1(1285)
\end{array}\right)=
\left(\begin{array}{cc} \cos\alpha&-\sin\alpha\\
\sin\alpha&\cos\alpha
\end{array}\right)
 \left(\begin{array}{l}
\mbox{\bf 8}\\
\mbox{\bf 1}
\end{array}\right),
\label{81}
\end{eqnarray}
where $\alpha$ is the $f_1(420)-f_1(1285)$ mixing angle in the  singlet-octet basis.

 With the help of
\begin{eqnarray}
\left(\begin{array}{l}
\mbox{\bf 8}\\
\mbox{\bf 1}
\end{array}\right)=
\left(\begin{array}{cc}
\sqrt{\frac{1}{3}}&-\sqrt{\frac{2}{3}}\\
\sqrt{\frac{2}{3}}&\sqrt{\frac{1}{3}}
\end{array}\right)
\left(\begin{array}{l}
N\\
S
\end{array}\right),
\label{ns-81}
\end{eqnarray}
from (\ref{NS}) and (\ref{81}), one can have
\begin{eqnarray}
U=\left(\begin{array}{cc}
\cos\alpha&-\sin\alpha\\
\sin\alpha&\cos\alpha
\end{array}\right)
\left(\begin{array}{cc}
\sqrt{\frac{1}{3}}&-\sqrt{\frac{2}{3}}\\
\sqrt{\frac{2}{3}}&\sqrt{\frac{1}{3}}
\end{array}\right).
\label{umatrix}
\end{eqnarray}

Based on (\ref{mix}), (\ref{dig}) and (\ref{umatrix}), the following relations can be obtained
\begin{eqnarray}
&M^2_{a_1(^3P_1)}+2\beta=(\sqrt{\frac{1}{3}}\cos\alpha-\sqrt{\frac{2}{3}}\sin\alpha)^2M^2_1+(\sqrt{\frac{2}{3}}\cos\alpha+\sqrt{\frac{1}{3}}\sin\alpha)^2M^2_2,
\label{rel1}\\
&\sqrt{2}\beta
X=(\sqrt{\frac{1}{3}}\cos\alpha-\sqrt{\frac{2}{3}}\sin\alpha)(\sqrt{\frac{2}{3}}\cos\alpha+\sqrt{\frac{1}{3}}\sin\alpha)(M^2_2-M^2_1),
\label{rel2}\\
 &2M^2_{K_1(^3P_1)}-M^2_{a_1(^3P_1)}+\beta
X^2=(\sqrt{\frac{1}{3}}\cos\alpha-\sqrt{\frac{2}{3}}\sin\alpha)^2M^2_2+(\sqrt{\frac{2}{3}}\cos\alpha+\sqrt{\frac{1}{3}}\sin\alpha)^2M^2_1.
\label{rel3}
\end{eqnarray}

 The constituent quark
mass ratio can be determined within the nonrelativistic constituent quark model(NRCQM). In
NRCQM\cite{nrqm,prd57}, the mass of a $q\bar{q}$ state with $L=0$, $M_{q\bar{q}}$ is given by
\begin{eqnarray}
M_{q\bar{q}}=m_q+m_{\bar{q}}+\Lambda\frac{\mbox{\bf s}_q\cdot\mbox{\bf s}_{\bar{q}}}{m_q
m_{\bar{q}}},
\end{eqnarray}
where $m$ and $\mbox{\bf s}$ are the constituent quark mass and spin, $\Lambda$ is a constant.
Since $\mbox{\bf s}_q\cdot\mbox{\bf s}_{\bar{q}}=-3/4$ for spin-0 mesons and $1/4$ for spin-1
mesons, in the SU(2) flavor symmetry limit, one can have
\begin{eqnarray}
X\equiv\frac{m_u}{m_s}=\frac{M_{\pi}+3M_{\rho}}{2M_K+6M_{K^\ast}-M_{\pi}-3M_{\rho}}=0.6298\pm
0.00068.
\end{eqnarray}

Taking $\alpha\simeq 50^\circ$ obtained from several independent analyses\cite{zphysc76} as
mentioned in section 1, $M_1=1426.3\pm 0.9$ MeV and $M_2=1281.8\pm 0.6$ MeV\cite{pdg}, from
relations (\ref{rel1})-(\ref{rel3}), we have\footnote{Here $\beta\simeq 108078.0\pm 834.788
~\mbox{MeV}^2$.}
\begin{eqnarray}
M_{K_1(^3P_1)}\simeq 1307.35\pm 0.63 ~\mbox{MeV},~~M_{a_1(^3P_1)}\simeq 1205.06\pm 0.92
~\mbox{MeV}. \label{k3p1}
\end{eqnarray}

The $K_1(^3P_1)$ and $K_1(^1P_1)$ can mix to produce the physical states $K_1(1400)$ and
$K_1(1270)$ and the mixing between $K_1(^3P_1)$ and $K_1(^1P_1)$ can be parameterized
as\cite{suzuki}
\begin{eqnarray}
\begin{array}{ll}
K_1(1400)=&K_1(^3P_1)\cos\theta_K-K_1(^1P_1)\sin\theta_K,\\
K_1(1270)=&K_1(^3P_1)\sin\theta_K+K_1(^1P_1)\cos\theta_K,
\end{array}
\end{eqnarray}
where $\theta_K$ denotes the $K_1(^3P_1)-K_1(^1P_1)$ mixing angle.  Without any assumption about
the origin of the $K_1(^3P_1)-K_1(^1P_1)$ mixing,
 the masses
of the $K_1(^3P_1)$ and $K_1(^1P_1)$ can be related to $M_{K_1(1400)}$ and $M_{K_1(1270)}$, the
masses of the $K_1(1400)$ and $K_1(1270)$, by the following relation phenomenologically,
\begin{eqnarray}
S \left(\begin{array}{cc}
M^2_{K_1(^3P_1)}& A\\
A&M^2_{K_1(^1P_1)}
\end{array}\right)
S^\dagger= \left(\begin{array}{cc}
M^2_{K_1(1400)}&0\\
0&M^2_{K_1(1270)}
\end{array}\right),
\label{kakb}
\end{eqnarray}
where $A$ denotes a parameter describing the $K_1(^3P_1)-K_1(^1P_1)$ mixing , and
\begin{eqnarray}
S=\left(\begin{array}{cc}
\cos\theta_K&-\sin\theta_K\\
\sin\theta_K&\cos\theta_K
\end{array}\right).
\nonumber
\end{eqnarray}
From (\ref{kakb}), one can have
\begin{eqnarray}
&&M^2_{K_1(^3P_1)}=M^2_{K_1(1400)}\cos^2\theta_K+M^2_{K_1(1270)}\sin^2\theta_K,
\label{kmass1}\\
&&M^2_{K_1(^1P_1)}=M^2_{K_1(1400)}\sin^2\theta_K+M^2_{K_1(1270)}\cos^2\theta_K,
\label{kmass2}\\
&&\cos(2\theta_K)=\frac{M^2_{K_1(^3P_1)}-M^2_{K_1(^1P_1)}}{M^2_{K_1(1400)}-M^2_{K_1(1270)}}.
\label{kmass3}
\end{eqnarray}
Inputting $M_{K_1(1400)}=1402\pm 7$ MeV, $M_{K_1(1270)}=1273\pm 7$ MeV\cite{pdg} and
$M_{K_1(^3P_1)}\simeq 1307.35\pm 0.63$ MeV shown in (\ref{k3p1}), from
(\ref{kmass1})-(\ref{kmass3}), we have
\begin{eqnarray}
M_{K_1(^1P_1)}\simeq 1370.03\pm 9.69~\mbox{MeV},~~|\theta_K|\simeq  (59.55\pm 2.81)^\circ.
\label{k1p1} \label{mixangle}
\end{eqnarray}

 Recently, based on the relations (\ref{kmass1})-(\ref{kmass3}) and
restricting to $0<\theta_K < 90^\circ$, Nardulli and Pham found\cite{pham}
\begin{eqnarray}
&&\mbox{[solution a]:}~~(M_{K_1(^1P_1)}, M_{K_1(^3P_1)})=(1310,
1367)~\mbox{MeV},~\mbox{for}~~\theta_K=32^\circ,
\nonumber\\
&&\mbox{[solution b]:}~~(M_{K_1(^1P_1)}, M_{K_1(^3P_1)})=(1367,
1310)~\mbox{MeV},~\mbox{for}~~\theta_K=58^\circ. \nonumber
\end{eqnarray}
Our predicted result that $(M_{K_1(^1P_1)}, M_{K_1(^3P_1)})\simeq (1370, 1307)$
 MeV and $|\theta_K|\simeq 59.55^\circ$ extracted from $\alpha\simeq 50^\circ$ is in excellent
agreement with the solution b given by\cite{pham}.

 Within the nonrelativistic constituent quark model, the results regarding
the masses of the $K_1(^1P_1)$ and $K_1(^3P_1)$,  $(M_{K_1(^1P_1)}, M_{K_1(^3P_1)})=(1368, 1306)$
MeV suggested by \cite{nrqm} and $(M_{K_1(^1P_1)}, M_{K_1(^3P_1)})=(1356, 1322)$ MeV suggested by
\cite{prd57}, are in good agreement with our predicted result. However, based on the following
relation employed by \cite{nrqm,prd57}
\begin{eqnarray}
\tan^2(2\theta_K)=\left(\frac{M^2_{K_1(^3P_1)}-M^2_{K_1(^1P_1)}}{M^2_{K_1(1400)}-M^2_{K_1(1270)}}\right)^2-1,
\label{tan}
\end{eqnarray}
the values of $\theta_K = (31\pm 4)^\circ$ given by\cite{nrqm} and $\theta_K=(37.3\pm 3.2)^\circ$
given by\cite{prd57} disagree with value of $|\theta_K|\simeq (59.55\pm 2.81) ^\circ$ given by the
present work.

Obviously, (\ref{tan}) is equivalent to (\ref{kmass3}), and will yield two solutions $|\theta_K|$
and $\frac{\pi}{2}-|\theta_K|$. Simultaneously considering the relations (\ref {kmass1}),
(\ref{kmass2}) and (\ref{tan}), in the presence of $M_{K_1(1400)}>M_{K_1(1270)}$, we can conclude
that if $M_{K_1(^3P_1)}< M_{K_1(^1P_1)}$, the $|\theta_K|$ would greater than $45^\circ$ . In fact,
relation (\ref{kmass3}) clearly indicates that in the presence of $M_{K_1(1400)}>M_{K_1(1270)}$,
the case $M_{K_1(^3P_1)}< M_{K_1(^1P_1)}$ must require $45^\circ <|\theta_K|< 90^\circ$.

In the framework of a covariant light-front quark model, the calculations performed by Cheng and
Chua \cite{prd69} for the exclusive radiative $B$ decays, $B\rightarrow K_1(1270)\gamma$,
$K_1(1400)\gamma$, show that the relative strength of $B\rightarrow K_1(1270)\gamma$ and
$B\rightarrow K_1(1270)\gamma$ rates is very sensitive to the sign of the $K_1(1270)-K_1(1400)$
mixing angle. For $\theta_K=\pm 58^\circ$, the following relation is predicted\cite{prd69}
\begin{eqnarray}
\frac{{\cal{B}}(B\rightarrow K_1(1270)\gamma)} {{\cal{B}}(B\rightarrow K_1(1270)\gamma)}=
\left\{\begin{array}{ll}
 10.1\pm 6.2&\mbox{for}~~\theta_K=+58^\circ,\\
0.02\pm 0.02&\mbox{for}~~\theta_K=-58^\circ.
\end{array}\right.\end{eqnarray}
Evidently, experimental measurement of the above ratio of branching fractions can be used to fix
the sign of the $K_1(^3P_1)-K_1(^1P_1)$ mixing angle. Recently,  the first measurement of the
branching ratio ${\cal{B}}$ for $B$ decay into $K_1(1270)\gamma$, together with an upper bound on
$K_1(1400)$, ${\cal{B}}( B^+\rightarrow K^+_1(1270)\gamma)=(4.28\pm 0.94\pm 0.43)\times 10^{-5}$,
 ${\cal{B}}( B^+\rightarrow
K^+_1(1400)\gamma)< 1.44 \times 10^{-5}$ is reported by Belle collaboration\cite{belle}. Based on
the measurements of Bell collaboration\cite{belle}, the analysis of the radiative $B$ decays with
an axial-vector meson in the final state  performed by Nardulli and Pham\cite{pham} within naive
factorization suggests that ${\cal{B}}(B^+\rightarrow K^+_1(1400)\gamma)=4.4\times 10^{-6}$ for
$\theta_K=+58^\circ$, which is consistent with the predictions given by\cite{prd69}. Further
experimental studies of ${\cal{B}}( B^+\rightarrow K^+_1(1270)\gamma)$ and ${\cal{B}}(
B^+\rightarrow K^+_1(1400)\gamma)$ is certainly desirable for understanding the sign of the
$K_1(^3P_1)-K_1(^1P_1)$ mixing angle.

 Our predicted center value of the $a_1(^3P_1)$ mass is $\sim 1205.06$ MeV, slightly smaller than
the measured center value of the $a_1(1260)$ mass, $1230$ MeV, although the predicted value
$1205.06\pm 0.92$ MeV is consistent with the experimental datum $1230\pm 40$ MeV within errors. The
similar result has been obtained by Chliapnikov within NRCQM\cite{nrqm}. According to the NRCQM
prediction that if $M_{K_1(^3P_1)}<M_{K_1(^1P_1)}$, $M_{a_1(^3P_1)}$ would be less than
$M_{b_1(^1P_1)}$ \cite{nrqm,prd57}, therefore, in the presence of $M_{K_1(^3P_1)}\simeq 1307 <
M_{K_1(^1P_1)}\simeq 1370$ MeV, the $a_1({^3P_1})$ mass should smaller than the  $b_1(1230)$ mass
($1229.5\pm 3.2$ MeV\cite{pdg}).  In addition, notice that the determination of the $a_1(1260)$
mass in hadronic production and in $\tau\rightarrow a_1\nu_\tau$ decay is to a certain extent model
dependent\cite{pdg}.

\section{The $s\bar{s}$ member of the $^1P_1$ meson nonet}
\indent \vspace*{-1cm}

According to PDG\cite{pdg}, the $h_1(1170)$ as the $^1P_1$ isoscalar state (mostly of
$u\bar{u}+d\bar{d}$) is well established experimentally. However, the assignment of $s\bar{s}$
partner of the $h_1(1170)$ remains ambiguous. In the presence of the $b_1(1235)$ and $h_1(1170)$
being the members of the $^1P_1$ meson nonet, with the help of the $K_1(^1P_1)$ mass obtained in
section 2, we shall estimate the mass of the $^1P_1$ $s\bar{s}$ state using different approaches.

By applying (\ref{mix}) and (\ref{dig}) to the $^1P_1$ meson nonet, we can obtain the following
relations
\begin{equation}
\begin{array}{l}
2M^2_{K_1(^1P_1)}+(2+X^2)\beta_1=M^2_{h_1(1170)}+M^2_{h^\prime_1},\\
(M^2_{b_1(1235)}+2\beta_1)(2M^2_{K_1(^1P_1)}-M^2_{b_1(1235)}+\beta_1 X^2)-2\beta^2_1
X^2=M^2_{h_1(1170)}M^2_{h^\prime_1},
\end{array}
\label{trace}
\end{equation}
where $h^\prime_1$ denotes the $s\bar{s}$ partner of the $^1P_1$ states $h_1(1170)$ and $b_1(1235)$
.
 Using
$M_{K_1(^1P_1)}\simeq 1370.03\pm 9.69$ MeV, $X=0.6298\pm 0.00068$ obtained in section 2, and the
measured values $M_{b_1(1235)}=1229.5\pm 3.2$ MeV and $M_{h_1(1170)}=1170\pm 20$ MeV\cite{pdg}, we
have
\begin{eqnarray}
\beta_1\simeq -(69143.5\pm 22373.6) ~\mbox{MeV}^2,~~M_{h^\prime_1}\simeq 1489.75\pm
18.08~\mbox{MeV}.
 \label{h1x}
\end{eqnarray}
Then from (\ref{mix}) and (\ref{dig}), the quarkonia content of the $h_1(1170)$ and
$h^\prime_1(1490)$ can be given by
\begin{eqnarray}
\left(\begin{array}{l}
h^\prime_1(1490)\\
h_1(1170)
\end{array}\right)\simeq\left(\begin{array}{ll}
0.073\pm 0.02&-(0.997\pm 0.002)\\
0.997\pm 0.002&0.073\pm 0.02
\end{array}\right)\left(\begin{array}{ll}
N\\
S
\end{array}\right).
\label{content}
\end{eqnarray}

(\ref{h1x}) and (\ref{content}) indicate that with the $b_1(1230)$, $h_1(1170)$ and $K_1(1370)$ in
the $^1P_1$ meson nonet, another isoscalar state of the $^1P_1$ meson nonet, $h^\prime_1$ would
have a mass about 1490 MeV and is composed mostly of $s\bar{s}$.

Considering the fact that the $f^\prime_2(1525)$ is an almost pure $s\bar{s}$ state\cite{jpg27}, we
obtain the estimated mass of the $^1P_1$ $s\bar{s}$ state from the following relation given by
NRCQM\cite{nrqm}
\begin{eqnarray}
M_{s\bar{s}(^1P_1)}=M_{f^\prime_2(1525)}-(M_{a_2(1320)}-M_{b_1(1235)})X^2=1489.78\pm 5.16~
\mbox{MeV},
\end{eqnarray}
which is in excellent agreement with $M_{h^\prime_1}\simeq 1489.75\pm 18.08$ MeV shown in
(\ref{h1x}).

Also, in the framework of the quasi-linear Regge trajectory (see Ref.\cite{epjc37} and references
therein), i.e.,
\begin{equation}
J=\alpha_{i\bar{i^\prime}}(0)+\alpha^\prime_{i\bar{i^\prime}} M^2_{i\bar{i^\prime}},
\label{trajectory}
\end{equation}
where $i$ ($\bar{i^\prime}$) refers to the quark (antiquark) flavor, $J$ and $M_{i\bar{i^\prime}}$
are respectively the spin and mass of the $i\bar{i^\prime}$ meson, $\alpha_{i\bar{i^\prime}}(0)$
and $\alpha^\prime_{i\bar{i^\prime}}$ are respectively the intercept and slope of the trajectory on
which the  $i\bar{i^\prime}$ meson lies; For a meson multiplet, the parameters for different
flavors can be related by the following relations

(i) additivity of intercepts,
\begin{equation}
\alpha_{i\bar{i}}(0)+\alpha_{j\bar{j}}(0)=2\alpha_{j\bar{i}}(0), \label{intercept}
\end{equation}

(ii) additivity of inverse slopes,
\begin{equation}
\frac{1}{\alpha^\prime_{i\bar{i}}}+\frac{1}{\alpha^\prime_{j\bar{j}}}=\frac{2}{\alpha^\prime_{j\bar{i}}},
\label{slope}
\end{equation}
for the $^1P_1$ $q\bar{q}$ nonet, one can have\footnote{Here we take
$\alpha^\prime_{n\bar{n}}=0.7218, ~\alpha^\prime_{s\bar{s}}=0.6613$ and
$\alpha^\prime_{n\bar{s}}=0.6902$ GeV$^{-2}$\cite{epjc37}.}
\begin{eqnarray}
M_{s\bar{s}(^1P_1)}=\left
[\frac{2\alpha^\prime_{n\bar{s}}M^2_{K_1(^1P_1)}-\alpha^\prime_{n\bar{n}}M^2_{b_1(1235)}}{\alpha^\prime_{s\bar{s}}}\right
]^{\frac{1}{2}}=1506.01\pm 18.62~\mbox{MeV},
\end{eqnarray}
which is also consistent with $M_{h^\prime_1}\simeq 1489.75\pm 18.08$ MeV given in (\ref{h1x}).

In the presence of the $b_1(1235)$, $h_1(1170)$ and $K_1(^1P_1)$ (with a mass about 1370 MeV)
belonging to the $^1P_1$ meson nonet, the above three different and complementary approaches, i.e.,
meson-meson mixing, nonrelativistic constituent quark model and Regge phenomenology, consistently
suggest that the ninth member of the $^1P_1$ nonet has a mass about $1495.18\pm 8.82$ MeV (averaged
value of the above three predicted results) and is mainly strange. Our predicted mass of the
$^1P_1$ $s\bar{s}$ state is in good agreement with the values $1499\pm 16$ MeV suggested by
Chliapnikov in a nonrelativistic constituent quark model\cite{nrqm} and 1511 MeV recently found by
Vijande et al. in a constituent quark model\cite{vijande}.

Experimentally, the $h_1(1380)$ with $J^{PC}=1^{+-}$ was claimed to be observed in
$K\overline{K}\pi$ system by only two collaborations, LASS collaboration\cite{lass} (Mass: $1380\pm
20$ MeV, $\Gamma= 80\pm 30$ MeV) and Crystal Barrel collaboration\cite{cbarrel} (Mass: $1440\pm 60$
MeV, $\Gamma=170\pm 80$ MeV), and the observed decay mode of the $h_1(1380)$ (
$K\overline{K}^\ast$) favors the assignment of the $h_1(1380)$ as a $s\bar{s}$ state.

On the one hand, our predicted mass of the $^1P_1$ $s\bar{s}$ state, $1495.18\pm 8.82$ MeV, is
significantly larger than $1380\pm 20$ MeV. The prediction given by Godfrey and Isgur in a
relativized quark model\cite{isgod} for the mass of the $^1P_1$ $s\bar{s}$ state is 1.47 GeV, at
least 70 MeV higher than the measured result of LASS\cite{lass}. Therefore if the measured results
of LASS\cite{lass} are confirmed, the $h_1(1380)$ seems too light to be the $^1P_1$ $s\bar{s}$
member. The studies on the implications of large $N_c$ and chiral symmetry for the mass spectra of
meson resonances performed by Cirigliano et al.\cite{0305311} also disfavor the assignment of the
$h_1(1380)$ to $^1P_1$ $s\bar{s}$.

On the other hand, the predicted mass of the $^1P_1$ $s\bar{s}$ state is consistent with $1440\pm
60$ MeV within errors, and the calculations performed by Barnes et al.\cite{barnes} for the total
width of the $^1P_1$ $s\bar{s}$ state in the $^3P_0$ decay model also show that at this mass the
assignment of the $h_1(1380)$ as the $^1P_1$ $s\bar{s}$ state appears plausible. So if the measured
results of Crystal Barrel\cite{cbarrel} are confirmed, the $h_1(1380)$ would be a convincing
candidate for the $s\bar{s}$ partner of the $^1P_1$ state $h_1(1170)$.

Notice that the uncertainties of these measurements are rather large, and the $h_1(1380)$ state
still needs further confirmation\cite{pdg}. Without confirmed experimental information about the
$h_1(1380)$, the present results indicate that  the assignment of the $h_1(1380)$ as the $^1P_1$
$s\bar{s}$ member may be premature.

\section{Concluding remarks}
\indent \vspace*{-1cm}

 The studies on the implications of the $f_1(1285)-f_1(1420)$ mixing for the
$K_1(^3P_1)-K_1(^1P_1)$ mixing angle indicate that the $f_1(1285)-f_1(1420)$ mixing angle $\sim
50^\circ$ suggested by Close et al.\cite{zphysc76} implies that ($M_{K_1(^3P_1)}$,
$M_{K_1(^1P_1)}$)$\simeq$ (1307, 1370) MeV, which therefore suggests that the
$K_1(^3P_1)-K_1(^1P_1)$ mixing angle $\simeq \pm 59.55^\circ$. Experimental measurement of the
ratio of $B\rightarrow K_1(1270)\gamma$ and $B\rightarrow K_1(1270)\gamma$ rates  can be used to
fix the sign of the $K_1(^3P_1)-K_1(^1P_1)$ mixing angle. Also, with the $b_1(1235)$, $h_1(1170)$
and $K_1(^1P_1)$ in the $^1P_1$ meson nonet,  three different and complementary approaches, i.e.,
meson-meson mixing, nonrelativistic constituent quark model and Regge phenomenology, consistently
suggest that the $^1P_1$ $s\bar{s}$ member  has a mass about 1495.18 MeV.
 Our predicted mass of the $^1P_1$ $s\bar{s}$ state is significantly larger than the measured
value of the $h_1(1380)$ mass reported by LASS\cite{lass}, while consistent with that reported by
Crystal Barrel\cite{cbarrel}, which shows that without further confirmation on the $h_1(1380)$, the
assignment of the $h_1(1380)$ remains open.

 \noindent {\bf Acknowledgments:}
This work is supported in part by National Natural Science
Foundation of China under Contract No. 10205012, Henan Provincial
Science Foundation for Outstanding Young Scholar under Contract
No. 0412000300, Henan Provincial Natural Science Foundation under
Contract No. 0311010800, and Foundation of the Education
Department of Henan Province under Contract No. 2003140025.


\baselineskip 18pt
\begin{thebibliography}{99}
\bibitem{carnegie} R. K. Carnegie et al, Phys. Lett. B {\bf 68}, 289 (1977)
\bibitem{isgur}N. Isgur, M. B. Wise, Phys. Lett. B {\bf 232}, 113 (1989)
\bibitem{rosner} J. Rosner, Com. Nucl. Part. Phys. {\bf 16}, 109 (1986)
\bibitem{asner} D. M. Asner et al., Phys. Rev. D {\bf 62}, 072006 (2000)
\bibitem{cheng}H. Y. Cheng, Phys. Rev. D {\bf 67}, 094007 (2003)
\bibitem{suzuki} M. Suzuki, Phys. Rev. D {\bf 47}, 1252 (1997)
\bibitem{burakovksy} L. Burkovsky, T. Goldman,  Phys. Rev. D {\bf 56},1368 (1997)
\bibitem{nrqm}P. V. Chliapnikov, Phys. Lett. B {\bf 496}, 129 (2000)

\bibitem{prd57} L. Burakovsky, T. Goldman, Phys. Rev. D {\bf 57}, 2879 (1998)
\bibitem{blundell} H. G. Blundell, S. Godfrey, B. Phelps, Phys. Rev. D {\bf 53}, 3712 (1996)
\bibitem{barnes}T. Barnes, N. Black, P. R. Page, Phys. Rev. D {\bf 68}, 054014 (2003)
\bibitem{isgod}S. Godfrey, N. Isgur, Phys. Rev. D {\bf 32}, 189 (1985)
\bibitem{godkok} S. Godfrey, R. Kokoski, Phys. Rev. D {\bf 43}, 1679 (1991)
\bibitem{vijande} J. Vijande, F. Fernandez, A. Valcarce, J. Phys. G {\bf31}, 481 (2005)
\bibitem{pdg} S. Eidelman et al., Phys. Lett. B {\bf 592}, 1 (2004)
\bibitem{zphysc76} F. E. Close, A. Kirk, Z. Phys. C {\bf 76}, 469 (1997)
\bibitem{cpl17} D. M. Li, H. Yu, Q. X. Shen, Chin. Phys. Lett. {\bf 17}, 558 (2000)
\bibitem{jpa35} W. S. Carvalho, A. S. de Castro, A. C. B. Antunes, J. Phys. A {\bf 35}, 7585 (2002)

\bibitem{epjc37} D. M.  Li et al.,  Eur. Phys. J. C {\bf 37}, 323 (2004)

\bibitem{jpg27} D. M. Li, H. Yu, Q. X. Shen, J. Phys. G {\bf 27}, 807 (2001)

\bibitem{pham} G. Nardulli, T. N. Pham, hep-ph/0505048
\bibitem{prd69}H. Y. Cheng, C. K. Chua, Phys. Rev. D {\bf 69}, 094007 (2004)
\bibitem{belle}H. Yang et al., Belle Collaboration, Phys. Rev. Lett. {\bf 94}, 111802 (2005)
\bibitem{lass} D. Aston et al., LASS Collaboration, Phys. Lett. B {\bf 201}, 573, (1988)
\bibitem{cbarrel} A. Abel et al., Crystal Barrel Collaboration, Phys. Lett. B {\bf 415}, 280 (1997)
\bibitem{0305311} V. Cirigliano, G. Ecker, H. Neufeld, A. Pich, JHEP {\bf 0306}, 012 (2003)
\end{thebibliography}
\end{document}